\documentstyle[epsf,twocolumn,aps]{revtex}
\begin{document}
\title{Electronic Characteristics of \\
Quasi-2D Metallochloronitrides:  \\ Na$_x$HfNCl (T$_c$=25 K)} 

\author{R. Weht, A. Filippetti, and W. E. Pickett}
\address{Department of Physics, University of California,
Davis CA 95616}

\maketitle

\begin{abstract}
Local density functional results are presented for the electron-doped
metallochloronitrides {\it A}$_x$ZrNCl and 
{\it A}$_x$HfNCl, {\it A}=Li or Na,
which superconduct up to 25K.  The alkali non-stoichiometry is treated
in a virtual crystal approximation.  The electronic structure is
strongly two dimensional, especially in the conduction band region
occupied by the carriers, because the states are formed from the in-plane
orbitals $d_{xy}, d_{x^2-y^2}$ of the metal ion and the $p_x, p_y$
orbitals of the N ion.  We predict a change of behavior at a doping
level of $x$=0.3.
\end{abstract}

\section*{Introduction}
High temperature superconductivity (HTS) in the layered cuprates 
continues to puzzle even after a dozen years of intense scrutiny.  Even
outside of the class of HTSs, however, the appearance of high T$_c$ (in 
the pre-cuprates sense) also is baffling.  The highest T$_c$ is
achieved in the fullerides, with T$_c$=40 K reported,\cite{erwin}
which are three dimensional (3D)
materials with, however, a great deal of zero dimensional (cluster)
character.  T$_c \sim$ 35 K is achieved in Ba$_{1-x}$K$_x$BiO$_3$, also
a 3D system, with only some flat portions of Fermi surface to bring
in questions of low dimensionality.  There is not yet any clear
theoretical accounting for the magnitude of T$_c$ in these two 
systems\cite{gunnar,sergej} although some phonon modes in the
fullerides are strongly coupled to the conduction electrons.
The A15 superconductors (T$_c$ up to 23 K) present a class in which the
relatively high value of T$_c$ is understood in terms of
strong electron-phonon coupling and a high density of states (DOS)
[N(E$_F$] at the Fermi level E$_F$.\cite{a15} 

A new, 2D material has recently been added to the list of (non-HTS)
superconductors with unusually high T$_c$.  Yamanaka {\it et al.}\cite{zrncl} 
discovered superconductivity up to 12 K in Li-doped ZrNCl, and up
to 25.5 K in Na-doped HfNCl.\cite{hfncl}
Shamoto {\it et al.}
\cite{sham1,sham2} have reproduced the superconductivity and
determined the crystal structure of superconducitng materials using
synchrotron X-ray diffraction data.  We provide here the first calculation 
of the electronic structure of this system, considering in particular
variation with doping level.

\section*{Description of Calculations}
\subsection*{Electronic Structure Methods}
We have applied the local density approximation\cite{cepald}
using two methods:
the linearized augmented plane wave (LAPW) method\cite{wien} 
for the bands and density of
states (DOS) that we show, and the plane wave ultrasoft pseudopotential
method\cite{usp} for the phonon frequencies.   Both methods produced
essentially the same band structures.  

To account for the doping we have used a virtual crystal treatment.  For
example, for $x$ fractional occupancy of a site by Na (ten core electrons
and one valence electron), that site is fully occupied by a nucleus of
charge $10+x$ with the corresponding number of electrons.  As expected,
Na is ionized with the electrons going into bands that are primarily
metal $d$ bands, so the virtual crystal treatment should be reasonable.
We will find that it is important to treat this charge self-consistently,
because non-rigid-band behavior appears not far from the region of
immediate interest.

\subsection*{Structure}
We have used the structures of Shamoto {\it et al.}\cite{sham1,sham2}
for these materials.  Because a Cl-HfN-HfN-Cl slab is only weakly
coupled to its neighboring slabs, we expect that the stacking sequence
(for example, whether hexagonal or rhombohedral) is not important, and
indeed that was found to be the case.  One such structure is shown in
Fig. 1, and the slab structure is the same for Hf and Zr, with only the
stacking sequence varying with doping.~\cite{str}

\section*{Discussion of Results}
The band structures are strongly two dimensional, so we discuss only the
in-plane dispersion here.
The undoped compounds have a band gap of 1.7-1.8 eV, between valence
bands that are 2$p$ states of Cl and N, and conduction states that are
mostly Hf or Zr.  Thus the formal ionic description
(Hf,Zn)$^{4+}$N$^{3-}$Cl$^{-}$ is reasonable, but obscures the strong
metal-N hybridization.

\begin{figure}[tbp]
\epsfxsize=4.5cm\centerline{\epsffile{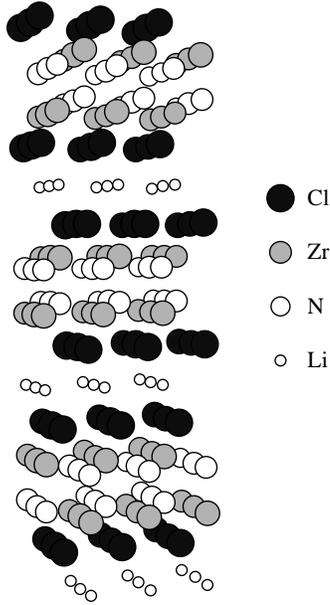}}
\vspace{10pt}
\caption{Rhombohedral structure of Li$_x$ZrNCl determined
by Shamoto {\it et al.}[7,8] 
The stacking along $c$ can change with doping,
but the basic slab structure structure is unchanged with doping.
Li sites are only fractionally occupied}
\label{f1:fig1}
\end{figure}

\begin{figure}[tbp]
\epsfxsize=7.0cm\centerline{\epsffile{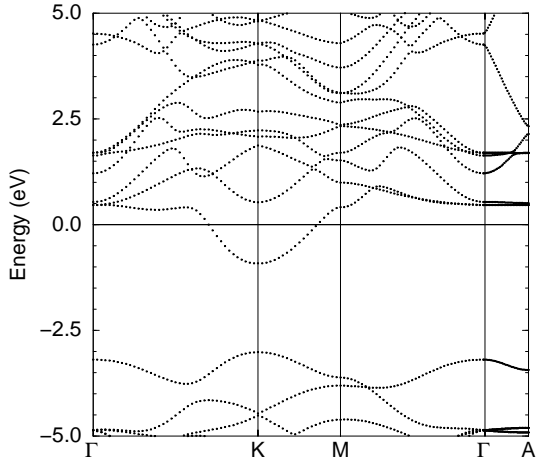}}
\vspace{10pt}
\caption{Plot along symmetry lines of the band structure of
Na$_{0.25}$HfNCl.  Points are $\Gamma=(0,0,0)$, $K= (2/3,1/3,0)$,
$M= (1/2,0,0)$, $A= (0,0,1/2)$, in units of the hexagonal reciprocal
lattice vectors.  The lack of dispersion along $\Gamma$-A of the lower 
conduction bands reflects the strong 2D character of the important
conduction band.}
\label{f2:fig2}
\end{figure}

\begin{figure}[tbp]
\epsfxsize=7.5cm\centerline{\epsffile{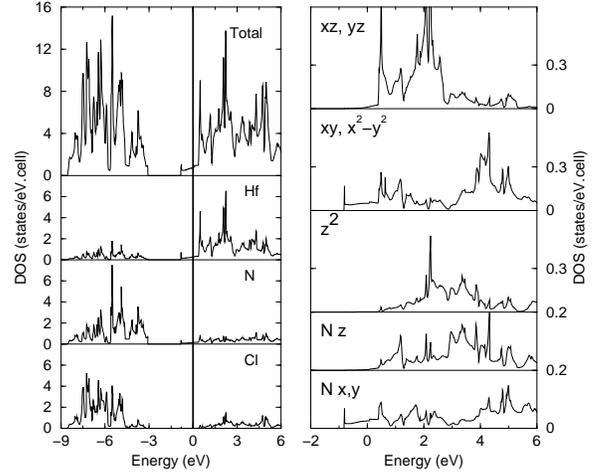}}
\vspace{10pt}
\caption{Atom (left panel) and orbital (right panel)
projected DOS for Na$_{0.25}$HfNCl.  In the right panel only the 
conduction bands are pictured.}
\label{f3:fig3}
\end{figure}

In Fig. 2 the band structure of Na$_{0.25}$HfNCl is shown along the
hexagonal symmetry directions.  Only a single band is occupied by the
doped carriers, and the band structure is very similar to that of the
undoped compound, {\it i.e.} a rigid band picture is good.
The projected DOS of Fig. 3 illustrates several things: (1) the occupied
band is a hybridized planar band of Hf $d_{xy}, d_{x^2-y^2}$ and N
$p_x, p_y$ character; (2) Hf $d_{xz}, d_{yz}$ and N $p_x$ character do
not appear until the band flattening 0.3 eV above E$_F$ in Fig. 2; (3) the
Hf $d_{z^2}$ character is higher still.  This behavior is
completely different from what Woodward and Vogt\cite{wood} calculated
for a bilayer structure reported by Juza and collaborators,\cite{juza}
in which two Hf ions are bonded across the bilayer,
rather than the Hf-N bonding across the bilayer as in the structure
determined by Shamoto.  Thus electrons are not doped into a very flat
band, which almost certainly would be strongly correlated, but rather into
a broad band with a light mass $m^* \approx 0.6 m$.

The other material parameters we obtain (for $x$=0.25) are \\
$\bullet$ Fermi velocity  v$_F = 3 \times 10^7$ cm/s\\
$\bullet$ Drude plasma energy  $\Omega_p= [4\pi e^2$ 
                     N(E$_F$)$v^2_{F,x}$]$^{1/2}$= 1.5 eV\\
$\bullet$ Gap 2$\Delta \approx$ 3.5 k$_B$ T$_c \approx$  7.5 meV \\
$\bullet$ Coherence length $\xi =\frac{\hbar v_F}{\pi \Delta}  \approx$ 180 \AA \\
$\bullet$ London penetration depth $\Lambda = \frac{c}{\Omega_p}$= 1300 \AA \\
$\bullet$ $\kappa =\Lambda/\xi  =  7 \rightarrow$ Type II superconductor.

\begin{figure}[tbp]
\epsfxsize=4.5cm\centerline{\epsffile{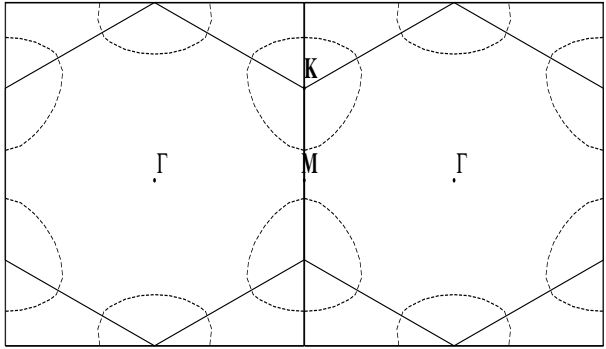}}
\caption{Plot of the distorted circular Fermi surface of Na$_{0.25}$HfNCl
shown in two hexagonal Brillouin zones.  The surface is centered at the
zone corner point K; since there are two such inequivalent points, there
are two Fermi surfaces.}
\label{f4:fig4}
\end{figure}

In Fig. 4 we present the Fermi surface for $x$=0.25 doping level.  It 
consists of threefold distorted circles at the zone corner (K) points.
There are two inequivalent such points, each with mean radius
$k_F = \frac{2}{9} \frac{\pi}{a}$.  On-Fermi-surface scattering
processes will be dominated by (1) small Q$\le 2k_F$ intraband scattering,
and (2) large Q$\sim$K interband scattering. [Note that the vector
connecting two neighboring zone corners K is also the vector K.]

We have calculated the frequencies of the three fully symmetric (A$_{1g}$)
Raman active vibrational modes of pristine ZrNCl and HfNCl
at the zone center, corresponding to modulation of the internal
coordinates $z_{Zr/Hf}, z_N, z_{Cl}$.  The frequencies are 586, 334, and
202 $cm^{-1}$ and 604, 368, and 191 $cm^{-1}$ respectively for the
two compounds.  The highest frequency is almost pure N motion. 

\section*{Doping}
As these systems are studied further, the question of the effect of doping
will be very important.  We have performed virtual crystal calculations
for several doping levels, with results for $x$ = 0.25, 0.35 and 0.45 
shown in Figure 5.  There are continuous rigid-band-like
changes up to just above $x$
= 0.30, above which bands with different character (specifically, out
of plane orbitals) begin to become occupied. At this point in doping the
properties should show a change in slope (plotted versus $x$).
At $x$=0.35 E$_F$ coincides
with a band that is very flat band along $\Gamma$-K and has a small mass
along $\Gamma$-M as well, {\it i.e.} a peak in the density of states.
This range of doping should prove very interesting, and whether T$_c$
rises or falls will reveal important characteristics of the pairing
mechanism.
\begin{figure}[tbp]
\epsfxsize=7.5cm\centerline{\epsffile{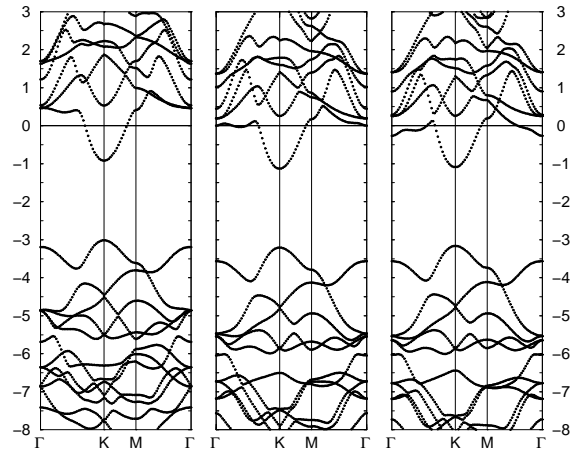}}
\vspace{10pt}
\caption{Band structure plots of Na$_x$HfNCl, for $x$=0.25, 0.35,
and 0.45 (left to right).  Around $x$=0.35 the Fermi level enters
a region where the bands have different character, and already the
conduction band dispersion in visibly non-rigid-like.  At $x$=0.45 there is a 
large Fermi surface centered at $\Gamma$ as well as the ones at K.
}
\label{f5:fig5}
\end{figure}

\section*{Summary and Acknowledgments}
It is surprising that such  low mass, low DOS materials such as those
described here can superconduct up to 25 K.  Since there is no indication,
experimentally or theoretically, of strong correlation effects in the
range of doping reported by the superconductors, a possible
candidate for pairing mechanism could be phononic.  Given the 2D
character and the strong nesting that might introduce characteristics of
1D behavior, 
purely electronic pairing mechanisms should also be considered.  Given the
apparent extreme two dimensionality of the crystal, it will be important
to establish that coherent supercurrent actually will flow perpendicular
to the layers.

We are indebted to R. Seshadri for many communications on the ZrNCl
system, to comments from D. J. Scalapino,
and to S. Shamoto for communication of unpublished work
including structural data.
This research was supported by Office of Naval Research
Grant No. N00014-97-1-0956 and National Science foundation Grant
DMR-9802076.


\begin{references}
\bibitem{str}For Na$_{0.29}$HfNCl the structure is
rhombohedral R$\overline 3$m (\#166), with $a$=3.5892~\AA, $c$=29.722 \AA.
Internal coordinates for the atoms (each in
$6c$ sites) are $z_{Hf}$=0.2083, $z_N$=0.1368, $z_{Cl}$=0.3944.  Na
was refined in the $3a$ site at the origin.  Li$_{0.16}$ZrNCl
is isomorphous, with $a$=3.60 \AA~being slightly larger.
\end{references}
\end{document}